%% file: HIN-16-024_temp.tex
\pdfoutput=1

\documentclass[11pt,twoside,a4paper,cmspaper,final,collab]{cms-tdr}
\def\svnVersion{481344P}\def\cmsCernNoTag{CERN-EP-2018-097}\def\cmsCernDate{\today}\def\cmsMessage{Published in the Journal of High Energy Physics as \href{http://dx.doi.org/10.1007/JHEP10(2018)161}{\doi{10.1007/JHEP10(2018)161.}}}

\begin{document}\cmsNoteHeader{HIN-16-024}

\hyphenation{had-ron-i-za-tion}
\hyphenation{cal-or-i-me-ter}
\hyphenation{de-vices}
\RCS$HeadURL: svn+ssh://svn.cern.ch/reps/tdr2/papers/HIN-16-024/trunk/HIN-16-024.tex $
\RCS$Id: HIN-16-024.tex 477215 2018-10-06 10:29:26Z alverson $

\newlength\cmsFigWidth
\setlength\cmsFigWidth{0.4\textwidth}
\providecommand{\cmsLeft}{left\xspace}
\providecommand{\cmsRight}{right\xspace}

\cmsNoteHeader{HIN-16-024}

\newcommand {\pp}            {\ensuremath{\Pp\Pp}\xspace}
\newcommand {\PbPb}          {\ensuremath{\mathrm{PbPb}}\xspace}
\newcommand {\ak}            {anti-\kt\xspace}
\newcommand {\DeltaRij}      {\ensuremath{\Delta R_{ij}}\xspace}
\newcommand {\DeltaRsj}      {\ensuremath{\Delta R_{12}}\xspace}

\providecommand{\ptjet}      {\ensuremath{\pt^\text{jet}}\xspace}
\providecommand{\ptg}        {\ensuremath{p_\mathrm{T,g}}\xspace}
\providecommand{\mg}         {\ensuremath{M_{\mathrm{g}}}\xspace}
\providecommand{\zg}         {\ensuremath{z_{\mathrm{g}}}\xspace}
\providecommand{\zcut}       {\ensuremath{z_{\text{cut}}}\xspace}
\providecommand{\mgptjet}    {\ensuremath{\mg/\ptjet}\xspace}

\newcommand {\SDA}           {$(0.1,0.0)$ SD setting\xspace}
\newcommand {\SDB}           {$(0.5,1.5)$ SD setting\xspace}

\newcommand {\subjet}        {subjet\xspace}
\newcommand {\subjets}       {subjets\xspace}

\newcommand {\softdrop}      {{soft drop}\xspace}

\newcommand {\mylumipp}      {\ensuremath{ 27.4\pbinv}\xspace}
\newcommand {\mylumipbpb}    {\ensuremath{ 404\mubinv}\xspace}

\newcommand {\JEWEL}         {{\textsc{jewel}}\xspace}
\newcommand {\QPYTHIA}       {{\textsc{q-pythia}}\xspace}

\title{Measurement of the groomed jet mass in \PbPb and \pp collisions at $\sqrtsNN = 5.02\TeV$}

\date{\today}

\abstract{
A measurement of the groomed jet mass in \PbPb and \pp collisions at a nucleon-nucleon center-of-mass energy of 5.02\TeV with the CMS detector at the LHC is presented.
Jet grooming is a recursive procedure which sequentially removes soft constituents of a jet until a pair of hard subjets is found.
The resulting groomed jets can be used to study modifications to the parton shower evolution in the presence of the hot and dense medium created in heavy ion collisions.
Predictions of groomed jet properties from the \PYTHIA and \HERWIGpp event generators agree with the measurements in \pp collisions.
When comparing the results from the most central \PbPb collisions to \pp data, a hint of an increase of jets with large jet mass is observed, which could originate from additional medium-induced radiation at a large angle from the jet axis.
However, no modification of the groomed mass of the core of the jet is observed for all \PbPb centrality classes.
The \PbPb results are also compared to predictions from the \JEWEL and \QPYTHIA event generators, which predict a large modification of the groomed mass not observed in the data.
}

\hypersetup{%
pdfauthor={CMS Collaboration},%
pdftitle={Measurement of groomed jet mass in PbPb and pp collisions at sqrt(sNN) = 5.02 TeV},%
pdfsubject={CMS},%
pdfkeywords={CMS, physics, heavy ion collisions, jet quenching, jet mass}}

\maketitle

\section{Introduction}

In heavy ion collisions, scattering processes with large momentum transfer $Q$ (of order 100\GeV or more) between the partonic constituents of the colliding nuclei occur early.
Energy loss experienced by these high-momentum partons (quarks or gluons) as a result of their interactions with the colored, hot and dense quantum chromodynamics (QCD) medium created in heavy ion collisions (the quark-gluon plasma, or QGP)~\cite{Gyulassy:1990ye,Baier:1994bd}, was first observed at BNL RHIC~\cite{Adler:2003qi,Adams:2003kv,Arsene:2003yk,Back:2003qr} and then at the CERN LHC~\cite{Chatrchyan:2011sx,Aad:2010bu,Aamodt:2010jd}.
Interactions of the outgoing partons with the QGP are also expected to modify the angular and momentum distributions of the parton shower relative to proton-proton (\pp) collisions.
It was shown at the LHC that there is a significant amount of energy carried by soft particles at large angles relative to the axes of the jets produced by outgoing partons~\cite{CMSHIN14010,CMSHIN14016}.

Parton interactions with the QGP can increase the gluon radiation probability of the propagating partons and can also lead to modifications of the momentum sharing between split partons, as well as the angular scale of the splitting~\cite{Vitev:2005yg,Renk:2009nz,Renk:2010zx,Armesto:2011ht,Majumder:2014gda}.
After a hard splitting, where both resulting partons carry a significant fraction of the original energy, the two energetic partons then evolve into separate sprays of particles within the jet.
By isolating these two hard-radiation sources, the interactions of the color charges of the medium with the two outgoing highly energetic partons can be studied.

Jet grooming algorithms~\cite{Ellis:2009me,Butterworth:2008iy,Krohn:2009th,Dasgupta:2013ihk,Larkoski:2014wba}
remove large-angle, soft radiation inside a jet, revealing the underlying hard structure via the identification of two \subjets. In \pp collisions this reflects the first hard splitting process.
The properties of these \subjets provide information about medium interactions of the two partons that originated in a hard splitting.
The hard structure of the jet is also expected to be sensitive to semihard medium-induced gluon radiation~\cite{Salgado:2003gb,Mehtar-Tani:2016aco}, modifications of the initial parton splitting~\cite{Chien:2016led}, and the medium
response~\cite{Milhano:2017nzm}. A modification in the distribution of the shared momentum fraction, \zg, defined as the energy of the sub-leading (in transverse momentum, \pt) \subjet over the sum of the two energies of the two \subjets, was previously studied in lead-lead (\PbPb) collisions~\cite{Sirunyan:2017bsd}. The opening angle of the parton splitting provides additional information about the nature of the modifications in the medium~\cite{Chien:2016led,Mehtar-Tani:2016aco}. This motivates studies of the groomed jet mass (\mg), defined as the invariant mass
of the system consisting of the two \subjets, which is sensitive to both the parton splitting function and the opening angle between the two outgoing partons.  This measurement complements studies of the mass of the full jet without using grooming algorithms~\cite{AliceMass}, which makes such studies mostly sensitive to soft wide angle radiation.

In this paper, a measurement of the ratio of the groomed jet mass and the jet \pt in both \pp and \PbPb collisions using the \softdrop (SD) jet grooming algorithm~\cite{Larkoski:2014wba} with two parameter settings is presented.
This analysis uses \pp and \PbPb collision datasets corresponding to integrated luminosities of \mylumipp and \mylumipbpb, respectively, collected with the CMS detector~\cite{Chatrchyan:2008zzk} at the LHC in 2015 at a nucleon-nucleon center-of-mass energy of 5.02\TeV.

\section{The CMS apparatus and event selection}
\label{sec:samples}

The central feature of the CMS apparatus is a superconducting solenoid of 6\unit{m} internal diameter,
providing an axial magnetic field of 3.8\unit{T}. Within the solenoid volume are a silicon pixel
and strip tracker, a lead tungstate crystal electromagnetic calorimeter, and a brass and
scintillator hadron calorimeter, each composed of a barrel and two endcap sections.
A hadron forward (HF) calorimeter, covering the pseudorapidity range $3 < \abs{\eta} < 5$, complements the coverage provided by the barrel and endcap
detectors. Muons are detected in gas-ionization chambers embedded in the steel flux-return
yoke outside the solenoid. The first level of the CMS trigger system~\cite{CMSTrigger}, composed of specialized
hardware processors, uses information from the calorimeters and muon detectors to select the
most interesting events in a fixed time interval of less than 4\mus.
The high-level trigger processor
farm further decreases the event rate from around 100\unit{kHz} to 1 (2)\unit{kHz} for \pp (\PbPb) collisions before data storage.
A more detailed description of the CMS detector, together with a definition of the coordinate system used and the relevant kinematic variables, can be found in Ref.~\cite{Chatrchyan:2008zzk}.

Events with multiple collisions (pileup) within a bunch crossing have a negligible effect on the measurement, since the average number of additional collisions is less than 0.9 in both data sets, and much lower in the \PbPb data set.
Events are selected with triggers requiring a jet with high \pt, found using the \ak algorithm~\cite{Cacciari2011, Cacciari:2008gp} with a distance parameter of $R = 0.4$. In \pp collisions, these triggers are based on jets reconstructed from particle-flow (PF) candidates~\cite{PFPaper}.
An unprescaled trigger with a \ptjet threshold of 80\GeV is used.
In \PbPb collisions, triggers are based on jets reconstructed from calorimeter deposits including a subtraction for the uncorrelated underlying event (UE)~\cite{Kodolova2007}.
Triggers with multiple thresholds are employed to ensure that their efficiency is high for the full range of phase space considered in the analysis.
The thresholds for these triggers are ${\ptjet=60}$, 80 and 100\GeV.  The triggers with lower \ptjet thresholds are prescaled.

	Several offline event selections are applied to reject events from beam-gas, beam-pipe, beam halo, cosmic ray muons, and beam scraping interactions~\cite{Khachatryan:2016odn}.
   A requirement of a coincidence of three towers with at least 3\GeV of total transverse energy in the HF detectors on each side of the interaction point~\cite{Chatrchyan:2008zzk} is employed to reject purely electromagnetic interaction events between Pb nuclei.
In \pp collisions this coincidence requirement is not present, as the contamination from electromagnetic interactions is negligible. For both collision systems a requirement is placed on the primary vertex, the reconstructed vertex with the highest amount of activity, to be within 15\unit{cm} from the nominal interaction point along the beam direction and within 0.15\unit{cm} in the transverse plane.

   In order to cope with the high particle multiplicity \PbPb environment, the event reconstruction algorithms are modified compared to the ones used for \pp data.
   Although not identical between the two colliding systems~\cite{Khachatryan:2016odn}, the tracking efficiency is comparable within a few percent in the \pt range relevant to the analysis, and it is well modeled by simulation.
        The collision centrality for \PbPb events is determined using the total sum of transverse energy from the calorimeter towers in the HF region. The transverse energy distribution is used to divide the event sample into bins of percentage of the total hadronic interaction cross section~\cite{Chatrchyan:2011sx}.
In this analysis, we present the results in four event centrality classes: 0--10\%, 10--30\%, 30--50\%,  and 50--80\%, with 0\% being the most central collision, and four \ptjet ranges: 140--160, 160--180, 180--200, and 200--300\GeV.

The \PYTHIA6.246~\cite{Sjostrand:2006za} (tune Z2*~\cite{Chatrchyan:2013gfi}) event generator prediction is compared with experimental \pp data and used to study systematic effects.
For \PbPb collision simulation, events generated with \PYTHIA are embedded into an UE produced with the \HYDJET 1.9 event generator~\cite{Lokhtin:2005px}.  All generated events undergo a full \GEANTfour~\cite{geant} simulation of the CMS detector response. Additional samples for cross checks and for comparison with the data are produced with  \HERWIGpp 2.7.1~\cite{Bahr:2008pv} (tune EE5C~\cite{Seymour:2013qka}).

Predictions for medium-modified jets are generated with \JEWEL 2.2.0~\cite{Jewel} (both with and without recoil, \ie, the scattered recoiling particles from the medium) and \QPYTHIA 1.0.3~\cite{QPythia} where the PQM model~\cite{PQMModel} is used to model the medium.  In order to model the effect of the uncorrelated UE, the samples generated with \JEWEL and \QPYTHIA are embedded in a simulated thermal background with particle momenta following a Maxwell-Boltzmann
distribution~\cite{BoltzmannUE} with an average \pt of 1.2\GeV and an average energy density
corresponding to that from events in the 0--10\% centrality class
in \PbPb data.

\section{Jet reconstruction}

Offline particle candidates are reconstructed with the PF algorithm. This algorithm aims to reconstruct and identify each individual particle (PF candidate) using an optimized combination of
information from various elements of the CMS detector. For this analysis, the PF candidates are treated as massless.
Jets are clustered from PF candidates using the \ak algorithm with a distance parameter of 0.4.  Only jets with $\ptjet > 140$\GeV and ${\abs{\eta_{\text{jet}}}<1.3}$ are included in the analysis due to the trigger.

In \PbPb collisions, the constituents of the jet are corrected for the UE contribution using the ``constituent subtraction'' algorithm~\cite{Berta:2014eza}. This algorithm uses a particle-level approach that removes or corrects jet constituents for the uncorrelated background based on the average UE density in a given $\eta$ region. This particle-by-particle subtraction allows the correction of both the four-momentum of the jet and its substructure. A more detailed description of this method can be found in Ref.~\cite{Sirunyan:2017bsd}.

The energy of reconstructed jets is corrected to the particle level with the corrections derived from simulation and applied to the reconstructed jets in \pp and \PbPb collisions.
Additional corrections for the mismodeling of the detector response are also applied~\cite{Chatrchyan:2011ds,Khachatryan:2016kdb}.

\section{Groomed jet mass}
\label{sec:ana}

Jet grooming isolates the hard sub-components of a jet and removes soft and wide-angle radiation, thereby highlighting jet substructure features.
This procedure can be used to isolate a hard splitting in the parton shower evolution. The soft components of a jet can originate from many sources, including uncorrelated UE, initial state radiation, other uncorrelated hard scattering in the collision, or soft gluons radiated by the hard parton which initiated the jet.
The SD jet grooming algorithm is used to extract the hard structure of jets, which is sensitive to the impact of parton-medium interactions during the jet evolution.
With this grooming technique, the hard and soft parts of the jets can be separated in a completely theoretically controlled way~\cite{Dasgupta:2001sh,Dasgupta:2013ihk,Larkoski:2014wba,Dasgupta:2013via,Dasgupta:2015yua,Larkoski:2015zka}.
The procedure starts with a jet and
reclusters the constituents with the Cambridge--Aachen algorithm~\cite{Dokshitzer:1997in} to
form an angular-ordered structure.
A recursive pairwise declustering step is then performed. In each step during the grooming procedure, the softer leg of the considered \subjet pair is dropped if the SD condition is not satisfied, resulting in a smaller groomed \pt than that of the original jet.  The SD condition is the following~\cite{Larkoski:2014wba}:
\begin{linenomath}
\begin{equation}
   \zg = \frac{\min(p_{\mathrm{T},i},p_{\mathrm{T},j})}{p_{\mathrm{T},i}+p_{\mathrm{T},j}} > \zcut\left( \frac{\DeltaRij}{R_{0}} \right)^{\beta},
\end{equation}
\end{linenomath}
where the subscripts ``$i$'' and ``$j$'' indicate the \subjets at that step of the declustering, \DeltaRij is the distance between the two \subjets in the $\eta-\phi$ plane, $R_{0}$ is the jet resolution parameter, and \zcut and $\beta$ are adjustable parameters.  The parameter \zcut is the threshold for \zg when the two \subjets are separated by the jet resolution parameter $R_0$, and $\beta$ controls the grooming profile as a function of \subjet separation \DeltaRij.
When $\beta=0$, the SD grooming threshold is independent of \DeltaRij, and the grooming procedure is equivalent to the modified mass--drop tagger~\cite{Dasgupta:2013ihk}.
  The jet is discarded if the SD condition is never satisfied before only one constituent remains.  This constitutes less than 1\% of the jets for the grooming parameter settings used in this analysis.
  Once the SD condition is satisfied, the two \subjets at that position in the angular-ordered tree are used to compute the mass.
 Assuming that these last two constituents surviving the grooming procedure are massless, the groomed jet mass (\mg) is calculated from their energies and opening angle.
The main variable used in this analysis is the groomed jet mass divided by the ungroomed jet transverse momentum, \mgptjet.
For this observable, the characteristic Sudakov peak (caused by the evolution of the shower) stays the same as \ptjet is varied~\cite{Dasgupta:2013ihk}, which allows the study for modification on mass without convoluting with the \ptjet spectrum.

In this analysis, two sets of parameters are considered: $\zcut = 0.1$ with $\beta = 0.0$, denoted as \SDA, and $\zcut = 0.5$ with $\beta=1.5$, denoted as \SDB.
The first parameter set has the advantage of being largely insensitive to higher-order QCD corrections, such as multiple emissions ~\cite{Dasgupta:2013ihk,Dasgupta:2013via}, while the second one is preferred experimentally
since it reduces the impact from UE fluctuations by applying a stronger SD constraint for \subjets with larger opening angle,
thereby focusing on the core of the jet.

\begin{figure}[h!t]
   \centering
   \includegraphics[width=0.49\linewidth]{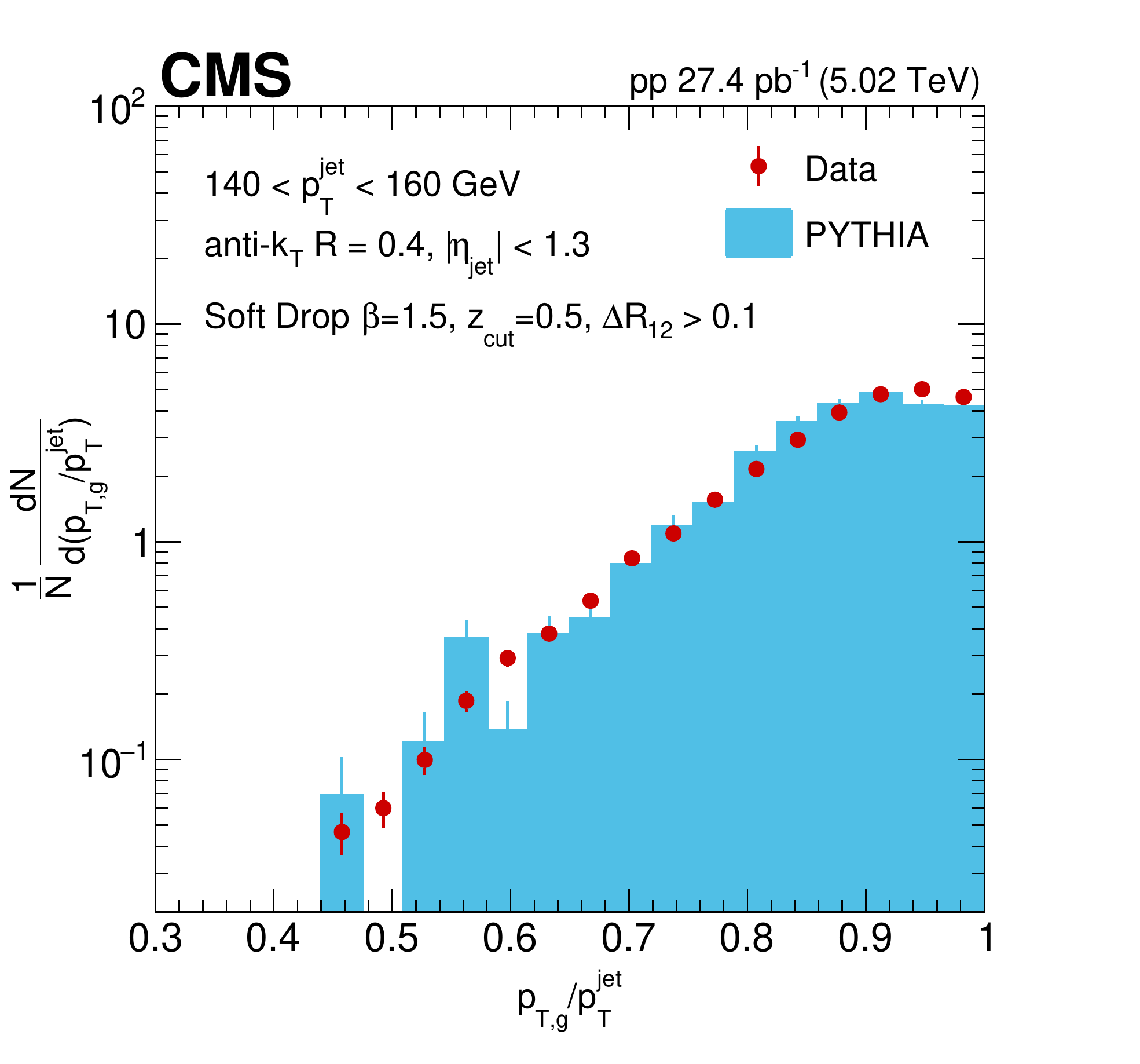}
   \includegraphics[width=0.49\linewidth]{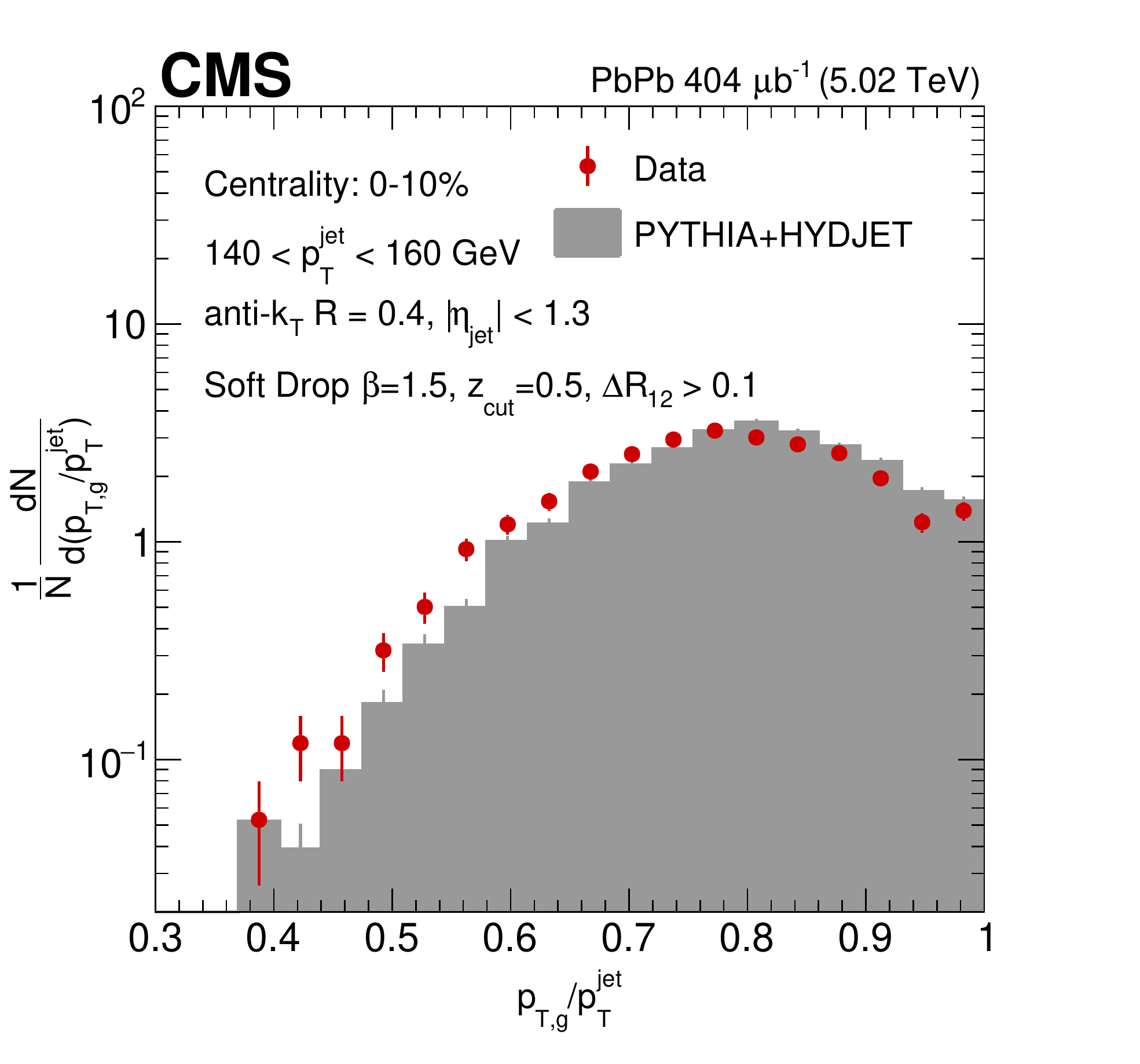}
   \caption{Groomed jet momentum fraction \ptg in \pp (\cmsLeft) and the 10\% most central \PbPb collisions (\cmsRight) for jets with $140 < \ptjet < 160\GeV$ and $\abs{\eta_{\text{jet}}} < 1.3$. The \pp data are compared to simulation using the \PYTHIA event generator and the \PbPb data are compared to the same \PYTHIA events embedded in \PbPb events simulated with the \HYDJET event generator.  Vertical lines indicate size of statistical uncertainty.
The parameters used for the SD algorithm are $\zcut = 0.5$, $\beta = 1.5$. The jets are selected based on the ungroomed jet transverse momentum.}
       \label{fig:GroomedEnergy}
\end{figure}
If two \subjets are very close to each other in the $\eta-\phi$ plane, they cannot be distinctly resolved, leading to a significant  worsening of the mass resolution. To avoid unphysical modification of the \mgptjet measurement, an additional selection on the \subjet opening angle of $\DeltaRsj > 0.1$ is applied.
For the 0--10\% \PbPb centrality bin, this \DeltaRsj requirement results in the rejection of 30\% of the jets using the \SDA and 50\% for the \SDB, due to a worse \subjet angular separation resolution when the UE is larger.  Both fractions are well reproduced by the simulation.

The groomed jet transverse momentum \ptg, divided by the ungroomed \ptjet in data, is compared to simulation at the reconstruction level in Fig.~\ref{fig:GroomedEnergy} for the \SDB.
More energy is removed in the 10\% most central \PbPb collisions than in \pp events in both data and simulation, indicating that the grooming procedure removes part of the residual background activity surviving the constituent subtraction procedure.
A difference in the $\ptg/\ptjet$ ratio distribution between data and simulation is seen in central \PbPb collisions due to correlated background, which is not modeled by the embedded sample.

Resolution effects in the \mgptjet distributions from charged-particle detection inefficiency, the particle angular resolution from the granularity of the calorimeter, and the UE fluctuations are not unfolded.
Instead, in order to compare results from \pp collisions with those of \PbPb collisions in a given \ptjet and centrality range, a smearing procedure is applied to the \pp data in order to account for the effects of the presence of the UE and differences in the reconstruction procedure between \PbPb and \pp data.
This is achieved by mixing a \pp event with a generated \PbPb UE at the reconstructed PF candidate level.  The UE is generated by sampling from the \pt spectra of the PF candidates in simulated minimum bias \PbPb events.
The PF candidates in the resulting mixed events are clustered and subtracted following the identical procedure used for the \PbPb data.
The ``smeared'' jets correspond to the expected modification in the presence of UE activity and detector effects but without any medium-induced modification to the jet structure.
The smearing procedure is validated using simulation by comparing with the embedded \PYTHIA + \HYDJET sample with full detector simulation with the smeared \PYTHIA sample.
In addition to the accounting for the resolution difference between \pp and \PbPb data, the smearing procedure also allows a better understanding of the different sources of systematic uncertainties.
The \mgptjet spectra in the PF-level embedding agrees within 3\% with that from the full detector simulation.
It is found that the dominant source causing this difference is the difference in tracking efficiency in \PbPb and \pp collisions.

The different track reconstruction in \PbPb and \pp collisions~\cite{Chatrchyan:2014fea,Khachatryan:2016odn} leads to a different \mg scale. A correction for \mgptjet is derived from simulation as a function of \DeltaRsj and applied to the smeared jets. The magnitude of the correction ranges from 1\% to 3\%, depending on the \subjet separation.  A good closure in the \mgptjet distribution between embedded and smeared jets is found.
The effect on \mgptjet from the merging of PF candidates  is found to be negligible compared to the \mg scale difference from the different tracking reconstruction algorithms.

\section{Systematic uncertainties}

The systematic uncertainties in the \mgptjet measurement are derived separately for \pp and \PbPb collisions.  Uncertainties are determined for each centrality and \ptjet selection. The following sources of systematic uncertainties are taken into account: online trigger, jet energy scale, jet energy resolution, \subjet angular resolution, smearing procedure, quark-to-gluon fraction, and the \mg scale correction.  Uncertainties in the UE associated with pileup collisions are found to be negligible
as compared to other uncertainties.

In \pp and \PbPb collisions with 30--100\% centrality, the trigger is fully efficient for jets in the kinematic range considered for this analysis.  For the 30\% most central \PbPb collisions, a trigger bias is present for the lowest considered \ptjet range, $140 < \ptjet < 160\GeV$. The measurement in this range is compared to the measurement using a lower-threshold trigger for which this effect is absent at $\ptjet=140$\GeV. The difference in the observed distributions, up to 5\% in the considered \mgptjet range, is
assigned as a systematic uncertainty.  It is also observed that the trigger used in the \pp data can induce a  bias to the smeared \mgptjet measurement for the 0--10\% central events in the lowest \ptjet bin.
As a result of the larger amount of smearing needed to compare to 0--10\% central events, a \pp jet with lower \ptjet where the trigger is not yet fully efficient may enter the analysis selection.
The  bias is studied by comparing the smeared jets collected with lower \ptjet threshold triggers.
An uncertainty of 7\% over the entire \mgptjet range is assigned.

The systematic uncertainty due to the jet energy scale (resolution) is estimated by changing the jet energy scale (resolution) by 5\% to cover the uncertainty on these quantities~\cite{Chatrchyan:2011ds}, followed by a comparison of the modified spectra with the nominal spectrum. The systematic uncertainty as a function of \mgptjet is derived from the difference between the spectra; it is generally of the order of 5\% for both jet energy scale and resolution.

The resolution of the opening angle between \subjets is found to be around 0.01
for a typical jet in this analysis with \subjet separation boundary of 0.1.  The effect of the angular resolution measurement on \mgptjet ratio is estimated by comparing spectra obtained by varying the selection on \DeltaRsj by 10\% up and down.  Only the low \mgptjet region is affected by changing the threshold, because of the correlation between \DeltaRsj and \mgptjet, resulting in an uncertainty
as large as 20\% for the \SDB.
Changes at high \mgptjet can be induced because the spectra are self-normalized.

Uncertainties associated with the \pp smearing procedure are obtained by varying the free parameters in the UE model.
The density of the UE is varied by 10\% which translates to a change in the \mgptjet spectrum by up to 10\% for $\mgptjet > 0.2$.
The fluctuation on the UE energy density is varied by 5\%, resulting in a change of the \mgptjet spectrum by 5\% across the entire range.

Since the fraction of quark- and gluon-initiated jets for a fixed \ptjet selection in \PbPb collisions is not known,
a systematic uncertainty is applied to the smeared jets in order to account for the different detector responses to quark and gluon jets.  It is estimated in simulation by taking half of the difference between smeared \mgptjet spectra for jets originated from quarks and gluons, and is found to be of order of 10--20\% towards the high tail ($\mgptjet > 0.2$).

The systematic uncertainty related to the \mg scale correction is estimated by comparing the smeared spectra obtained with different tracking algorithms used in \PbPb and \pp collisions data.  It is found that the change due to this is up to 6\% for larger values of \mgptjet and about 2\% in the bulk of the spectrum ($\mgptjet \simeq 0.05$--$0.10$).

\section{Results}

\begin{figure}[ht!]
   \centering
   \includegraphics[width=0.65\linewidth]{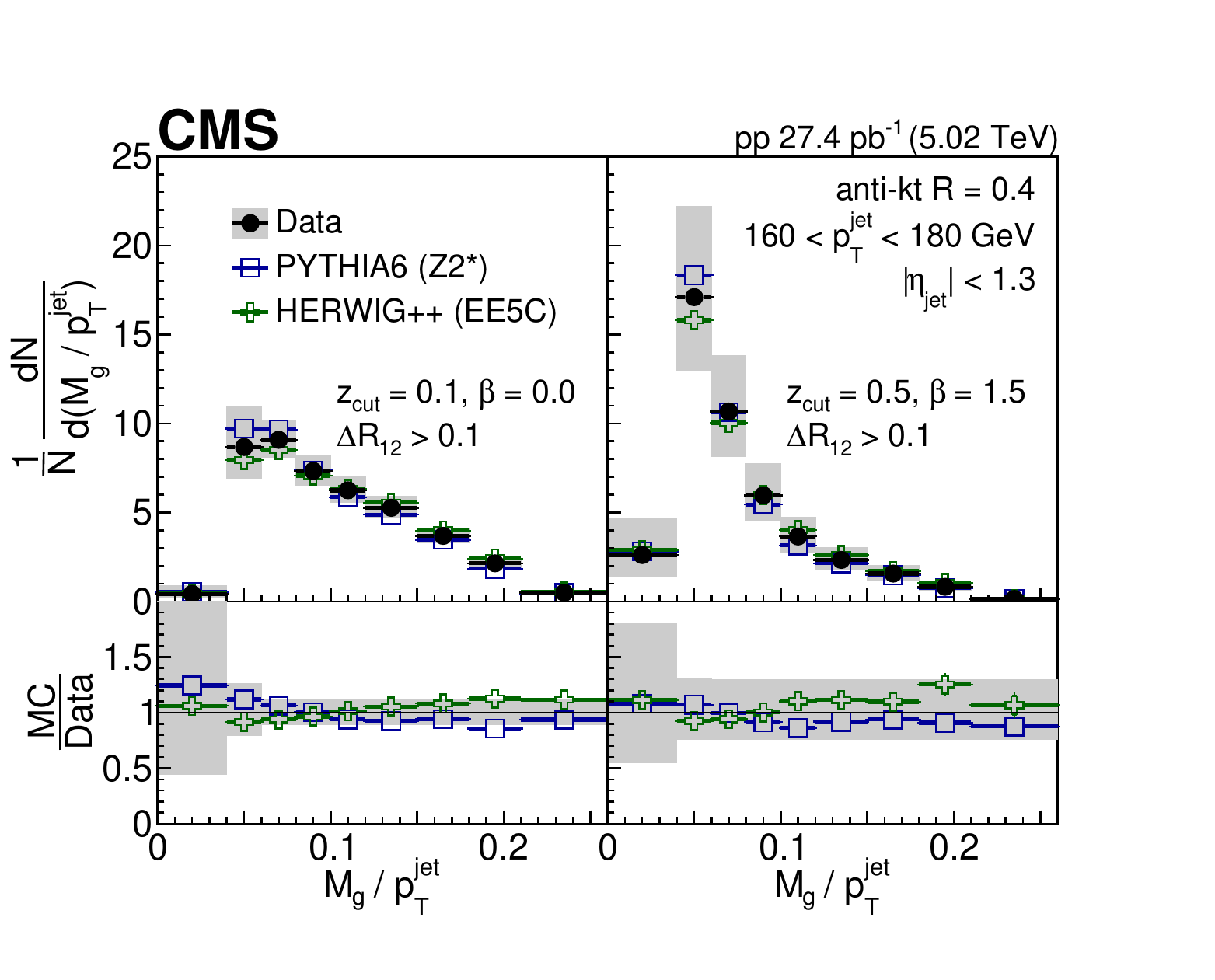}
   \caption{
      The spectra of \mgptjet for \pp events with $160 < \ptjet < 180\GeV$ using \SDA (left panels) and \SDB (right panels).  Results are compared to \PYTHIA and \HERWIGpp event generators.
The ratio of simulation to data is also shown.
   The heights of the gray boxes indicate systematic uncertainties.  Statistical uncertainties are less than the marker sizes.}
   \label{fig:PPResultSDA}
\end{figure}

The per jet normalized \mgptjet spectra in \pp collisions for various \ptjet selections are presented in Fig.~\ref{fig:PPResultSDA} for the $(0.1,0.0)$ and {\SDB}s. The results are compared to generated jets with \PYTHIA and \HERWIGpp.  At large \mgptjet, \HERWIGpp is above the \mgptjet spectra and \PYTHIA is below the spectra when compared to data with the \SDA, although the observed difference is smaller than the systematic uncertainties in the measurement.
The observed effect is in agreement with earlier measurements~\cite{ATLASJet1,ATLASJet2}.
A similar conclusion can be drawn for the \SDB.
With this setting, the \mgptjet spectrum is steeper than for the \SDA due to the larger amount of energy removed during the grooming procedure.
The lower edge of the spectra is caused by the \DeltaRsj requirement.

\begin{figure}[h!t]
  \centering
  \includegraphics[width=0.40\linewidth]{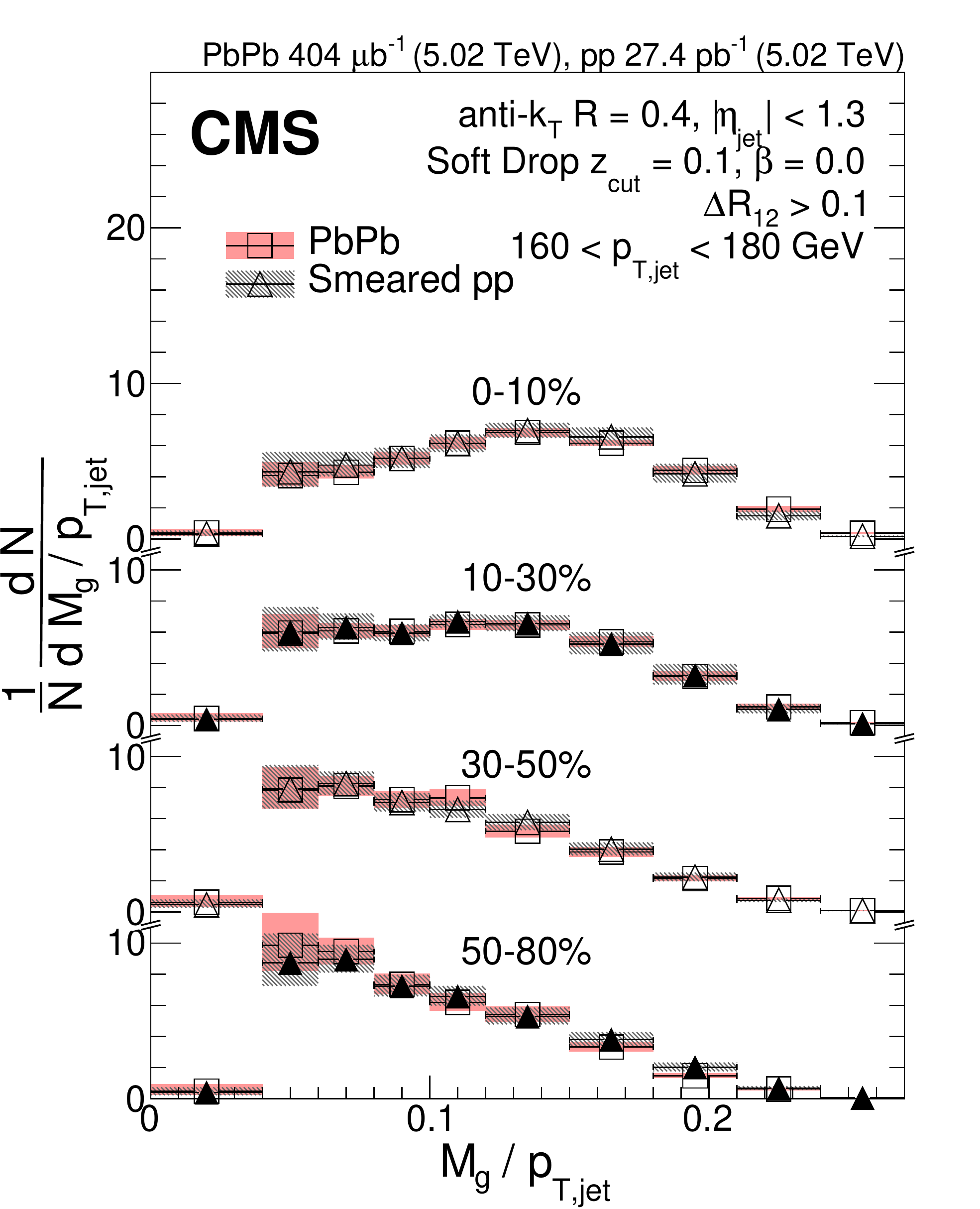}
  \includegraphics[width=0.40\linewidth]{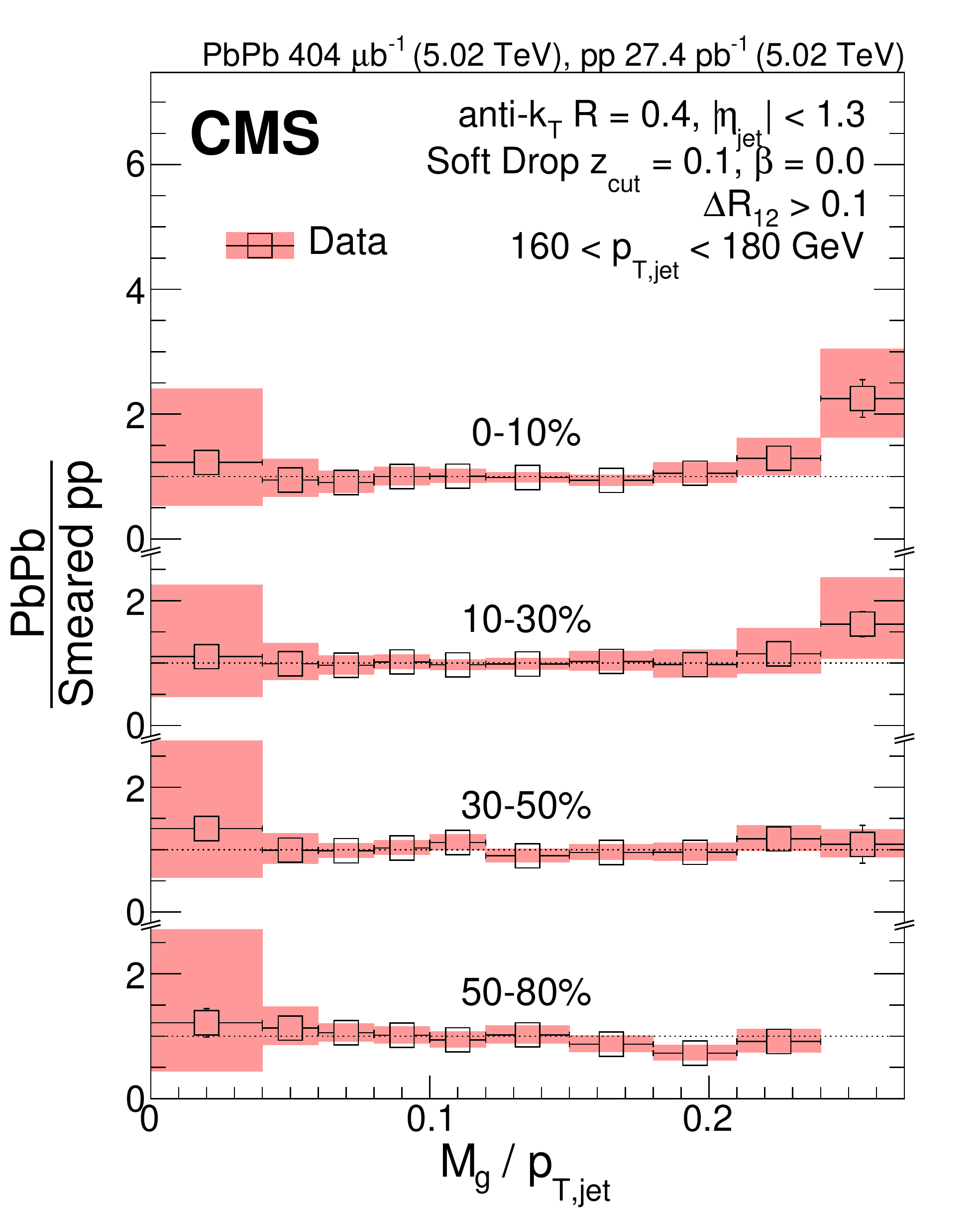}
   \caption{(\cmsLeft) The centrality dependence of \mgptjet, for \PbPb events with $160 < \ptjet < 180\GeV$ for the \SDA.  Results are compared  to the smeared \pp spectra.
   (\cmsRight) The ratio of \PbPb data over smeared \pp data.
   The heights of the vertical lines (colored boxes) indicate statistical (systematic) uncertainties.  Statistical uncertainties are less than the marker sizes in most bins.}
       \label{fig:SDMass0PbPbPPPtBin2}
\end{figure}

\begin{figure}[h!t]
  \centering
  \includegraphics[width=0.40\linewidth]{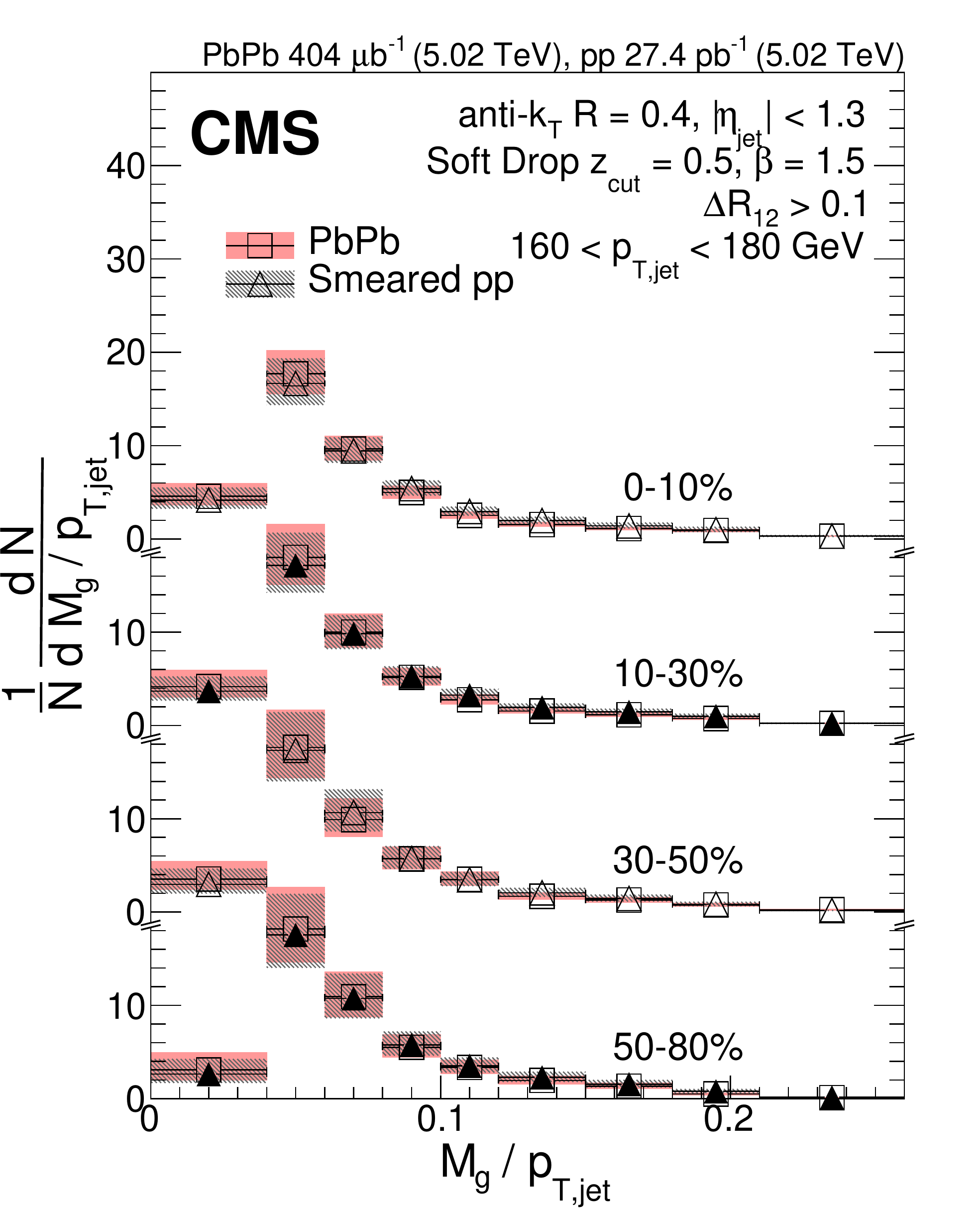}
  \includegraphics[width=0.40\linewidth]{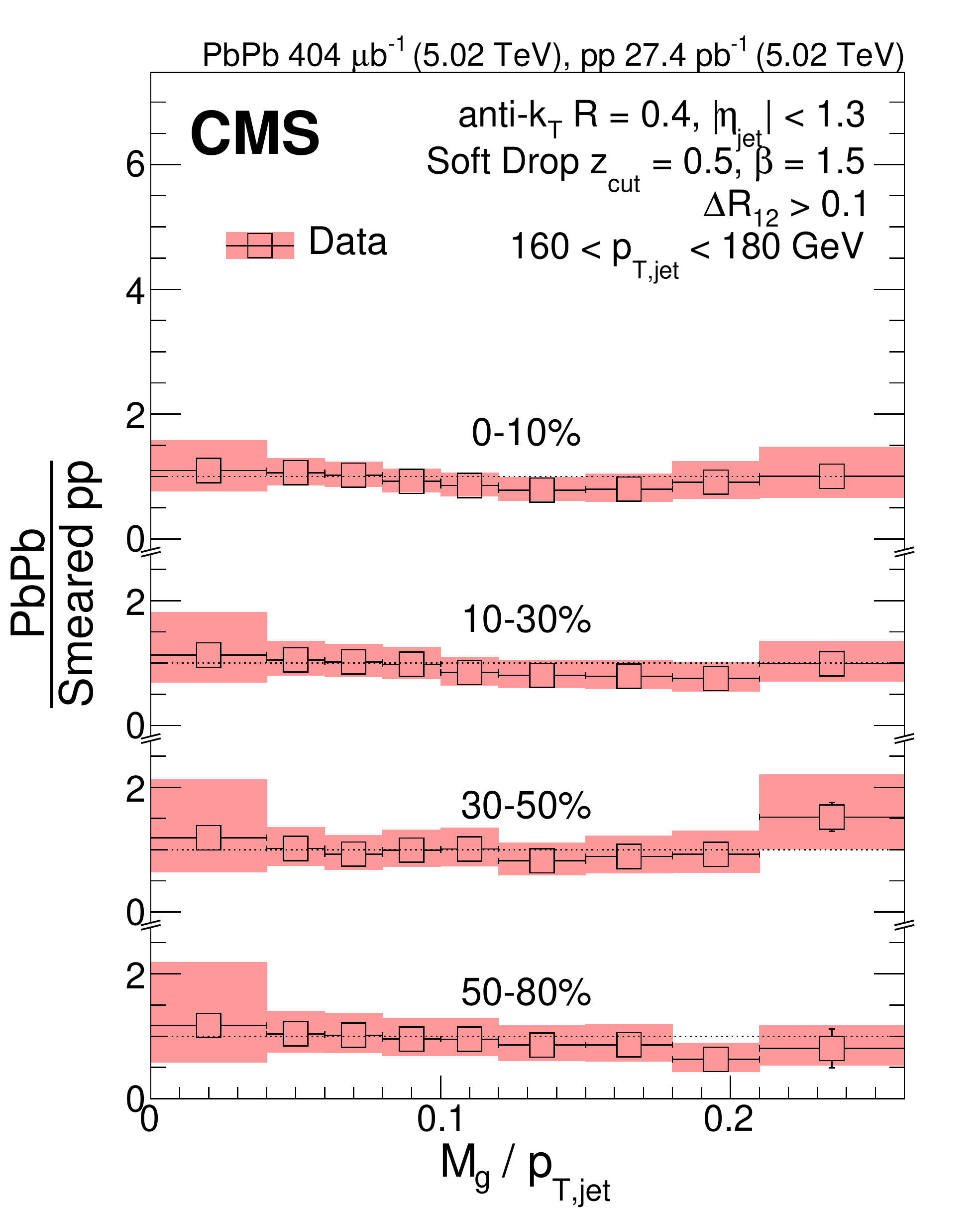}
   \caption{(\cmsLeft) The centrality dependence of \mgptjet, for \PbPb events with $160 < \ptjet < 180\GeV$ for the \SDB.  Results are compared  to the smeared \pp spectra.
   (\cmsRight) The ratio of \PbPb data over smeared \pp data.
   The heights of the vertical lines (colored boxes) indicate statistical (systematic) uncertainties.  Statistical uncertainties are less than the marker sizes in most bins.}
       \label{fig:SDMass7PbPbPPPtBin2}
\end{figure}

The measurement of the \mgptjet in \PbPb collisions for several centrality intervals for the \ptjet in the 160--180\GeV range is compared to the results for smeared \pp collisions in Figs.~\ref{fig:SDMass0PbPbPPPtBin2} and \ref{fig:SDMass7PbPbPPPtBin2} for the two SD grooming settings.
For the \SDA, no significant modification in \PbPb collisions compared to smeared \pp data is observed for this \ptjet range, except for a hint of an enhancement for the 10\% most central collisions.  For the \SDB, where the grooming disfavors pairs of \subjets with large opening angles and highly imbalanced \pt values, no noticeable modification is observed.
\begin{figure}[h!t]
  \centering
  \includegraphics[width=0.40\linewidth]{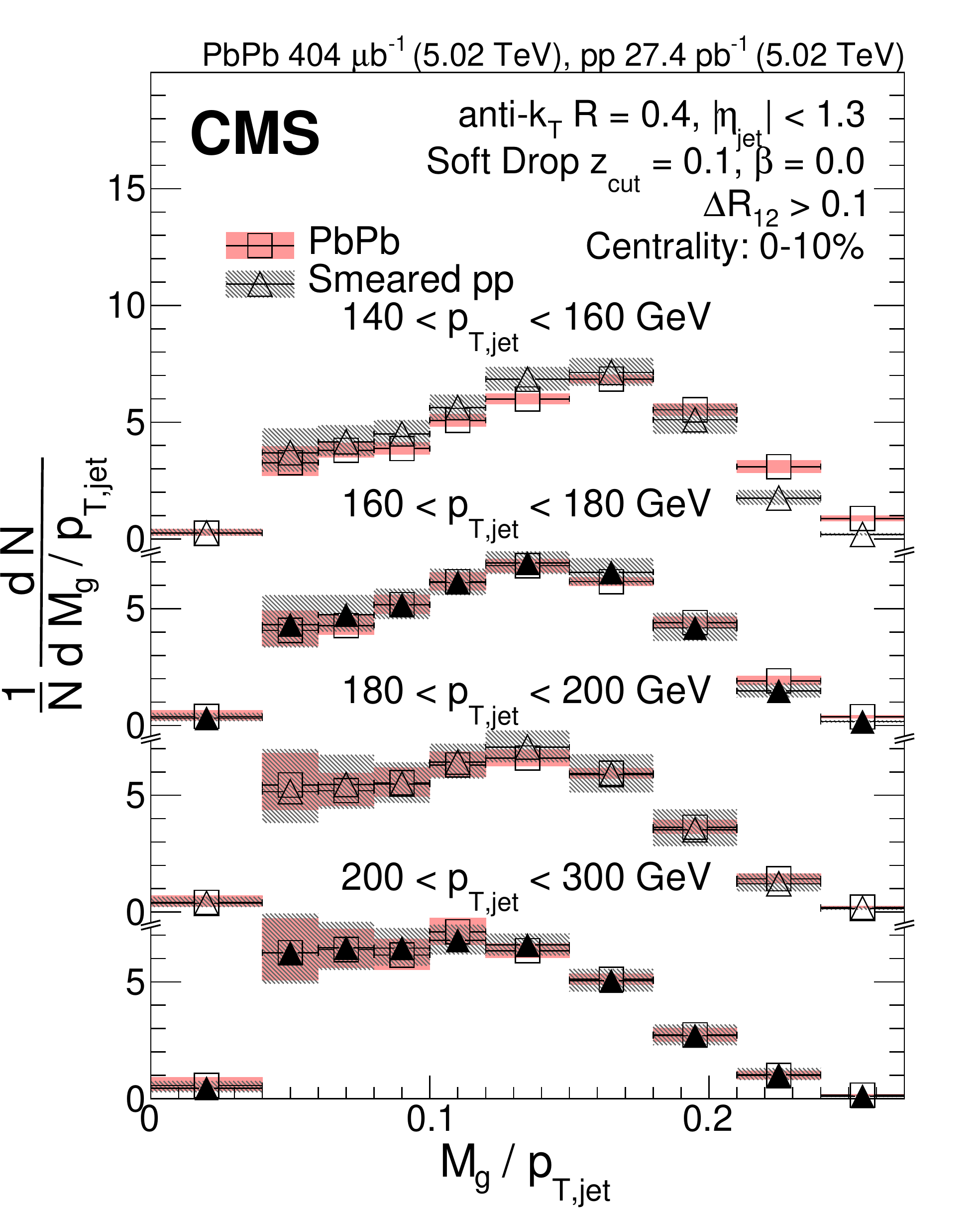}
  \includegraphics[width=0.40\linewidth]{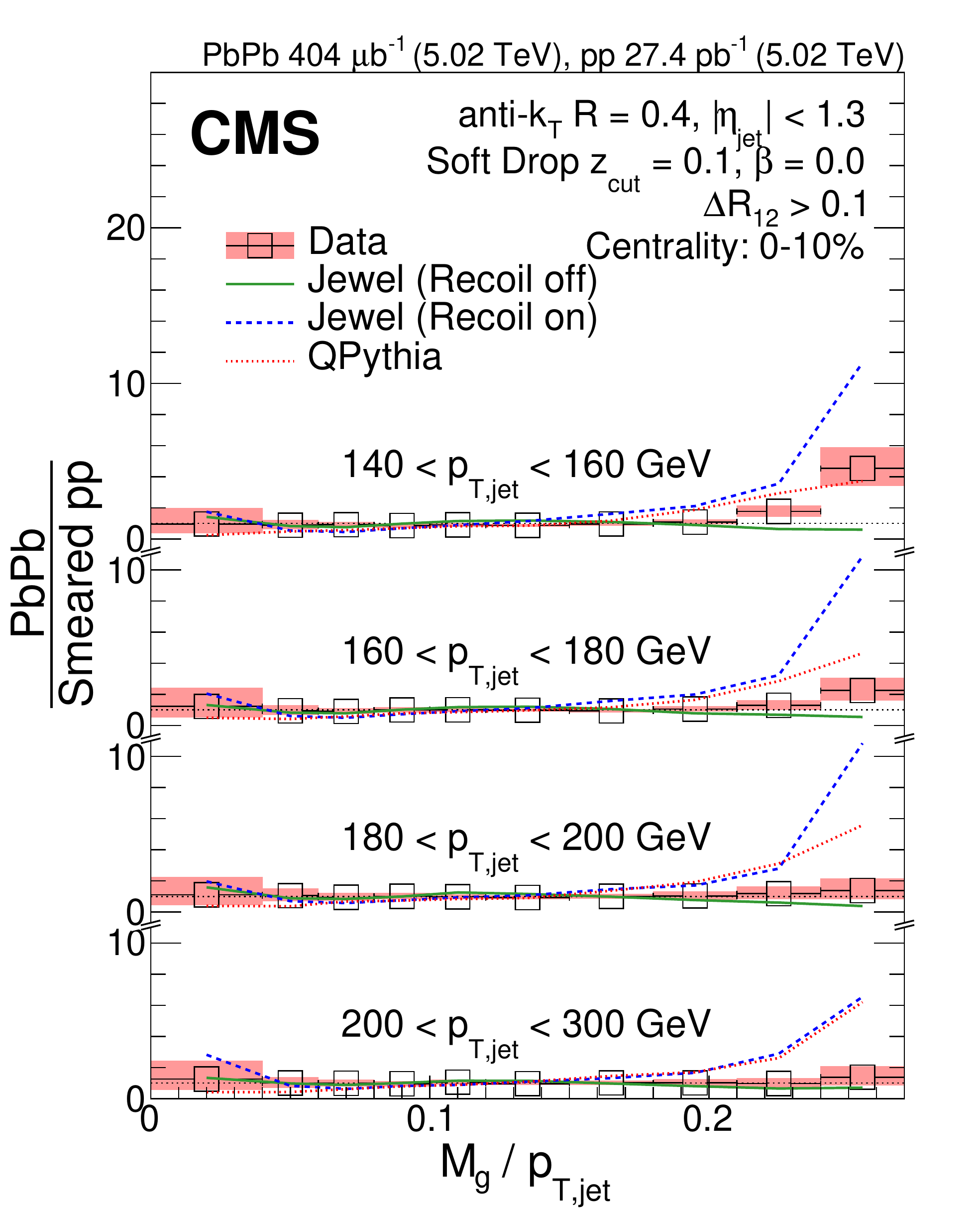}
   \caption{(\cmsLeft) The \ptjet dependence of \mgptjet, for \PbPb events in the centrality class 0--10\%, for the  \SDA.  Results are compared to the smeared \pp spectra.
   (\cmsRight) The ratio of \PbPb data over smeared \pp data.
The heights of the colored boxes indicate systematic uncertainties.  Statistical uncertainties are less than the marker sizes.  The ratios are compared to
smeared \JEWEL and \QPYTHIA generators, shown in blue and green, respectively.}
       \label{fig:SDMass0PbPbPPCentralityBin0}
\end{figure}
\begin{figure}[h!t]
  \centering
  \includegraphics[width=0.40\linewidth]{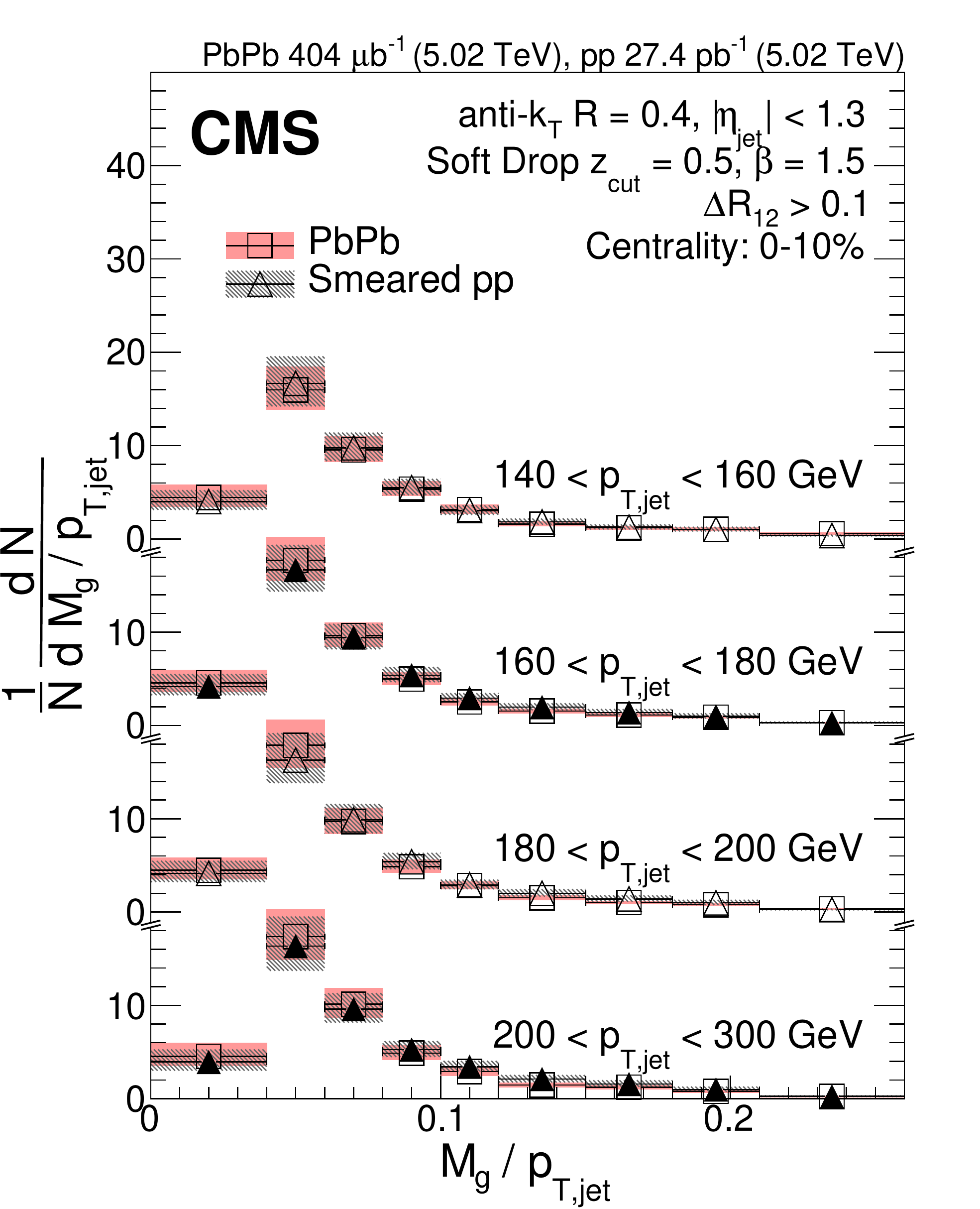}
  \includegraphics[width=0.40\linewidth]{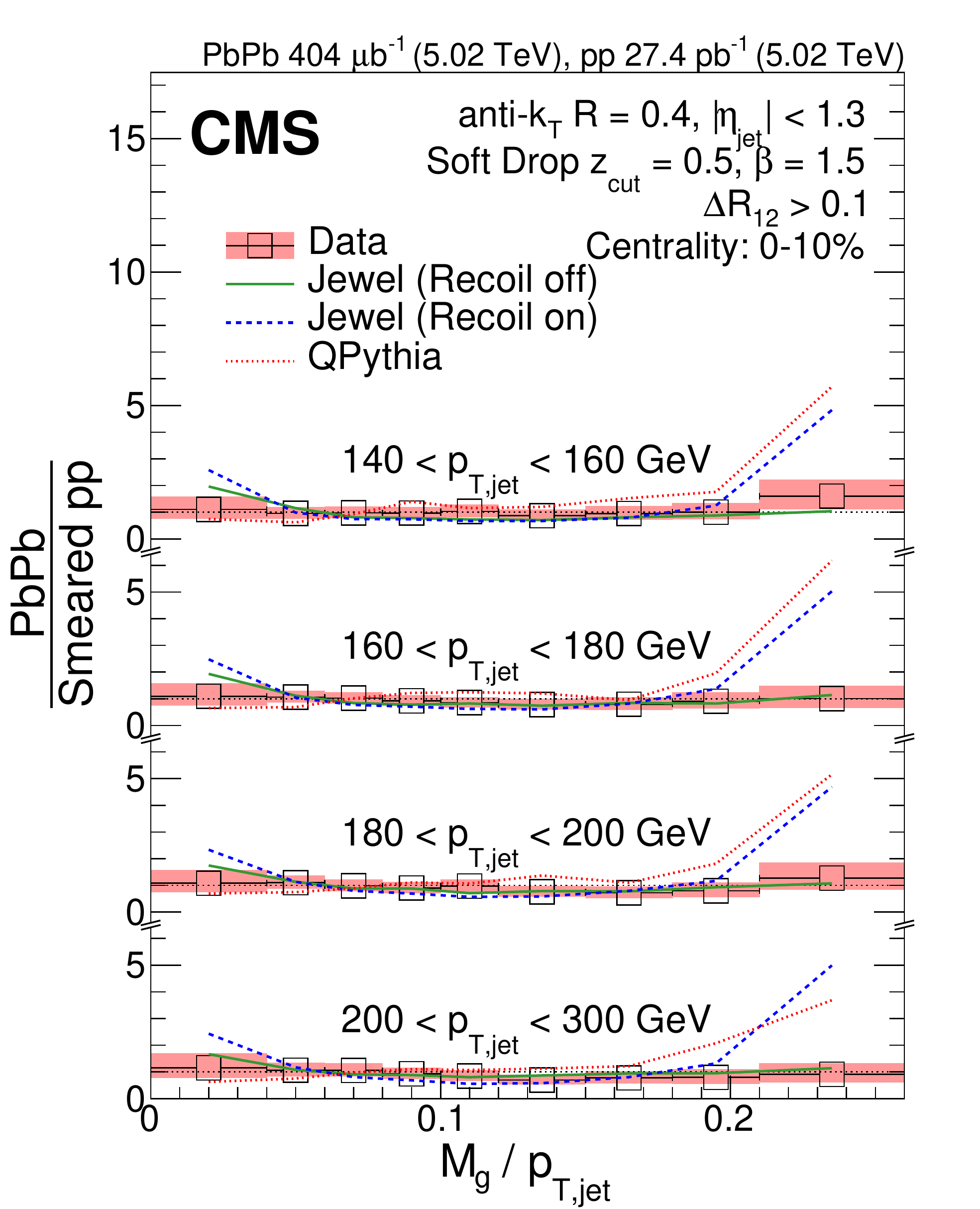}
   \caption{(\cmsLeft) The \ptjet dependence of \mgptjet, for \PbPb events in the centrality class 0--10\%, for the \SDB.  Results are compared to the smeared \pp spectra.
   (\cmsRight) The ratio of \PbPb data over smeared \pp data.
The heights of the colored boxes statistical (systematic) uncertainties.  Statistical uncertainties are less than the marker sizes.  The ratios are compared to
smeared \JEWEL and \QPYTHIA generators, shown in blue and green, respectively.}
       \label{fig:SDMass7PbPbPPCentralityBin0}
\end{figure}

In  Figs.~\ref{fig:SDMass0PbPbPPCentralityBin0} and \ref{fig:SDMass7PbPbPPCentralityBin0} the measured \mgptjet spectra in the 0--10\% \PbPb collisions sample are compared in several \ptjet intervals to the \pp smeared sample, for the two SD settings.
Some differences between jets from \PbPb collisions and smeared jets from \pp collisions are seen for the \SDA in the lowest \ptjet ranges.  This indicates that in central \PbPb collisions it is more likely to produce a jet with large \mgptjet than in \pp collisions. The
results are compared to two jet quenching event generators, which incorporate medium-induced radiation in the parton splitting process.  The generated events are smeared to account for effects from UE activity in \PbPb collisions.
The medium response in \JEWEL is modeled with the momentum transfers to recoiling scattering centers in the medium in addition to the splitting of jet constituents that is also present when the recoil feature in \JEWEL is disabled.
The relative enhancement of large-mass jets can be qualitatively captured by the \JEWEL generator with the recoil-on setting~\cite{Milhano:2017nzm,KunnawalkamElayavalli:2017hxo}, but the magnitude is much larger than that in data.  For the recoil-off setting, the enhancement at large \mgptjet is not reproduced, indicating that the recoil from the medium is important in reproducing the qualitative feature of the result.
In \QPYTHIA the medium modification enhances the splitting probability with an additional term that follows the BDMPS-Z radiation~\cite{QPythia,Baier:1996sk}.  This in turn increases the jet mass via the large amount of inter-jet broadening where the jets become less collimated.  The broadening of the mass distribution in \QPYTHIA is more prominent than in data.
The measured modifications are much smaller than predicted, as previously observed for the jet mass without grooming \cite{AliceMass}.

As a consequence of the stronger grooming at large \subjet opening angles, the result for the \SDB probes potential modification of the core of the jet.
On the contrary, in the \SDA the grooming strength does not depend on the \subjet opening angle and therefore is sensitive to both the core and peripheral modifications.
The comparison shows that the core of the jet is not altered in central \PbPb collisions within the uncertainties of the measurement, but the periphery of the jet is more sensitive to interactions of the partons with the dense colored medium during the parton shower evolution.
This effect vanishes at higher \ptjet and for more peripheral collisions.
The observed feature is not reproduced by theoretical models.
The comparison between the results from the two grooming settings indicates that the region of phase space included in the \SDA but excluded from the \SDB is the place with the most significant modification: splittings with large angular separation and low momentum sharing.

\section{Summary}

The first measurements of the ratio of the groomed jet mass and the transverse momentum of the jet, \mgptjet, in \pp and \PbPb collisions at a nucleon-nucleon center-of-mass energy of 5.02\TeV are presented. Both the \PYTHIA and \HERWIGpp event generators reproduce the measurement in \pp collisions.

{\tolerance=800
The results demonstrate that different grooming settings provide sensitivity to different parts of the phase space of \subjet angular separation and momentum sharing.
For \softdrop (SD) grooming parameters that remove more radiation at distances far away from the jet axis,
$({\zcut=0.5}, {\beta=1.5})$, the \mgptjet distribution in \PbPb collisions is, within uncertainties, in agreement with that measured in \pp collisions for all studied centrality (0--80\%) and \ptjet (140--300\GeV) regions.
Using the $({\zcut=0.1}, {\beta=0.0})$ SD setting, for which the grooming is independent of the angular separation of the \subjets, no significant modification of the \mgptjet spectra in 10--80\% peripheral collisions with respect to the measurement in \pp collisions is observed.
However, for the 10\% most central collisions, a hint of increased probability to produce jets with large \mgptjet is seen when compared to \pp collisions for jets with $140<\ptjet<180\GeV$.
The difference between the results from the two examined grooming settings indicates that the region of phase space where modifications are most significant are splittings with large angular separation and low-to-moderate momentum sharing.
The measurements are compared to the jet quenching event generators \JEWEL and \QPYTHIA, both of which predict a large enhancement at large \mgptjet that is not observed in the data.
\par}

\begin{acknowledgments}
We congratulate our colleagues in the CERN accelerator departments for the excellent performance of the LHC and thank the technical and administrative staffs at CERN and at other CMS institutes for their contributions to the success of the CMS effort. In addition, we gratefully acknowledge the computing centers and personnel of the Worldwide LHC Computing Grid for delivering so effectively the computing infrastructure essential to our analyses. Finally, we acknowledge the
enduring support for the construction and operation of the LHC and the CMS detector provided by the following funding agencies: BMWFW and FWF (Austria); FNRS and FWO (Belgium); CNPq, CAPES, FAPERJ, and FAPESP (Brazil); MES (Bulgaria); CERN; CAS, MoST, and NSFC (China); COLCIENCIAS (Colombia); MSES and CSF (Croatia); RPF (Cyprus); SENESCYT (Ecuador); MoER, ERC IUT, and ERDF (Estonia); Academy of Finland, MEC, and HIP (Finland); CEA and CNRS/IN2P3 (France); BMBF, DFG, and HGF
(Germany); GSRT (Greece); NKFIA (Hungary); DAE and DST (India); IPM (Iran); SFI (Ireland); INFN (Italy); MSIP and NRF (Republic of Korea); LAS (Lithuania); MOE and UM (Malaysia); BUAP, CINVESTAV, CONACYT, LNS, SEP, and UASLP-FAI (Mexico); MBIE (New Zealand); PAEC (Pakistan); MSHE and NSC (Poland); FCT (Portugal); JINR (Dubna); MON, RosAtom, RAS and RFBR (Russia); MESTD (Serbia); SEIDI, CPAN, PCTI and FEDER (Spain); Swiss Funding Agencies (Switzerland); MST (Taipei);
ThEPCenter, IPST, STAR, and NSTDA (Thailand); TUBITAK and TAEK (Turkey); NASU and SFFR (Ukraine); STFC (United Kingdom); DOE and NSF (USA).

\hyphenation{Rachada-pisek} Individuals have received support from the Marie-Curie programme and the European Research Council and Horizon 2020 Grant, contract No. 675440 (European Union); the Leventis Foundation; the A. P. Sloan Foundation; the Alexander von Humboldt Foundation; the Belgian Federal Science Policy Office; the Fonds pour la Formation \`a la Recherche dans l'Industrie et dans l'Agriculture (FRIA-Belgium); the Agentschap voor Innovatie door Wetenschap en Technologie (IWT-Belgium); the F.R.S.-FNRS and FWO (Belgium) under the ``Excellence of Science - EOS" - be.h project n. 30820817; the Ministry of Education, Youth and Sports (MEYS) of the Czech Republic; the Lend\"ulet (``Momentum") Programme and the J\'anos Bolyai Research Scholarship of the Hungarian Academy of Sciences, the New National Excellence Program \'UNKP, the NKFIA research grants 123842, 123959, 124845, 124850 and 125105 (Hungary); the Council of Science and Industrial Research, India; the HOMING PLUS programme of the Foundation for Polish Science, cofinanced from European Union, Regional Development Fund, the Mobility Plus programme of the Ministry of Science and Higher Education, the National Science Center (Poland), contracts Harmonia 2014/14/M/ST2/00428, Opus 2014/13/B/ST2/02543, 2014/15/B/ST2/03998, and 2015/19/B/ST2/02861, Sonata-bis 2012/07/E/ST2/01406; the National Priorities Research Program by Qatar National Research Fund; the Programa Estatal de Fomento de la Investigaci{\'o}n Cient{\'i}fica y T{\'e}cnica de Excelencia Mar\'{\i}a de Maeztu, grant MDM-2015-0509 and the Programa Severo Ochoa del Principado de Asturias; the Thalis and Aristeia programmes cofinanced by EU-ESF and the Greek NSRF; the Rachadapisek Sompot Fund for Postdoctoral Fellowship, Chulalongkorn University and the Chulalongkorn Academic into Its 2nd Century Project Advancement Project (Thailand); the Welch Foundation, contract C-1845; and the Weston Havens Foundation (USA).
\end{acknowledgments}

\bibliography{auto_generated}

\cleardoublepage \appendix\section{The CMS Collaboration \label{app:collab}}\begin{sloppypar}\hyphenpenalty=5000\widowpenalty=500\clubpenalty=5000\input{HIN-16-024-authorlist.tex}\end{sloppypar}
\end{document}

%% file: HIN-16-024-authorlist.tex
\vskip\cmsinstskip
\textbf{Yerevan Physics Institute, Yerevan, Armenia}\\*[0pt]
A.M.~Sirunyan, A.~Tumasyan
\vskip\cmsinstskip
\textbf{Institut f\"{u}r Hochenergiephysik, Wien, Austria}\\*[0pt]
W.~Adam, F.~Ambrogi, E.~Asilar, T.~Bergauer, J.~Brandstetter, M.~Dragicevic, J.~Er\"{o}, A.~Escalante~Del~Valle, M.~Flechl, R.~Fr\"{u}hwirth\cmsAuthorMark{1}, V.M.~Ghete, J.~Hrubec, M.~Jeitler\cmsAuthorMark{1}, N.~Krammer, I.~Kr\"{a}tschmer, D.~Liko, T.~Madlener, I.~Mikulec, N.~Rad, H.~Rohringer, J.~Schieck\cmsAuthorMark{1}, R.~Sch\"{o}fbeck, M.~Spanring, D.~Spitzbart, A.~Taurok, W.~Waltenberger, J.~Wittmann, C.-E.~Wulz\cmsAuthorMark{1}, M.~Zarucki
% ...
\vskip\cmsinstskip
% ...
1:  Also at Vienna University of Technology, Vienna, Austria\\
% ...
